\documentclass[prl,twocolumn]{revtex4-1}
\usepackage{amsmath}
\usepackage{amssymb}
\usepackage{array}
\usepackage{mathrsfs}
\usepackage{tabularx}
\usepackage{graphicx}
\usepackage{dcolumn}
\usepackage{bm}
\usepackage{graphicx}
\usepackage{color}
\usepackage{booktabs}
\usepackage{multirow}
\usepackage{diagbox}

\begin{document}
\title{280-km experimental demonstration of quantum digital signature with one decoy state}
\author{Hua-Jian Ding$^{1,2,3}$}
\author{Jing-Jing Chen$^{1,2,3}$}
\author{Liang Ji$^{1,2,3}$}
\author{Xing-Yu Zhou$^{1,2,3}$}
\author{Chun-Hui Zhang$^{1,2,3}$}
\author{Chun-Mei Zhang$^{1,2,3}$}
\author{Qin Wang$^{1,2,3}$}
\email{qinw@njupt.edu.cn}

\affiliation{$^{1}$Institute of quantum information and technology, Nanjing University of Posts and Telecommunications, Nanjing 210003, China.}
\affiliation{$^{2}$"Broadband Wireless Communication and Sensor Network Technology" Key Lab of Ministry of Education, NUPT, Nanjing 210003, China.}
\affiliation{$^{3}$"Telecommunication and Networks" National Engineering Research Center, NUPT, Nanjing 210003, China.}

\begin{abstract}
Quantum digital signature (QDS) guarantee the unforgeability, nonrepudiation and transferability of signature messages with information-theoretical security, and hence has attracted much attention recently. However, most previous implementations of QDS showed relatively low signature rates or/and short transmission distance. In this paper, we report a proof-of-principle phase-encoding QDS demonstration using only one decoy state. Firstly, such method avoids the modulation of vacuum state, thus reducing experimental complexity and random number consumption. Moreover, incorporating with low-loss asymmetric Mach-Zehnder interferometers and real-time polarization calibration technique, we have successfully achieved higher signature rate, e.g., 0.98 bit/s at 103 km, and to date a record-breaking transmission distance over 280-km installed fibers. Our work represents a significant step towards real-world applications of QDS.

\end{abstract}

\maketitle
\section{Introduction}
Digital signature \cite{DS}, as one basic primitive to ensure internet information security, is widely used in software distribution, financial transactions and electronic government, where the signed messages are unforgeable, nonrepudiated and transferable. Generally, classical digital signature are largely designed exploiting public-key cryptography, which relies on computational complexity, e.g. Rivest-Shamir-Adleman algorithm \cite{RSA} and elliptic-curve-based protocols \cite{EC1,EC2}. However, such schemes are potentially susceptible to the major breakthrough in algorithms and especially rapid development of quantum computers \cite{QC1,QC2}. Given the importance of digital signature, it is expected a more secure signature scheme to counter the eavesdropper (Eve) equipping with unlimited computational power. Quantum digital signature (QDS), analogously to quantum key distribution (QKD) \cite{BB84}, can offer information-theoretic security guaranteed by the fundamental laws of quantum mechanics. The first QDS was proposed by Gottesman and Chuang \cite{QDS-GC} in 2001. This scheme is enlightening, but currently unfeasible experimentally due to strict requirements of nondestructive state comparison, long-time quantum memory and a secure quantum channel. Thereafter, several experimentally friendly and high-throughput protocols have been introduced \cite{QDS-Andersson,QDS-Dunjko,QDS-Arrazola1,QDS-Wallden,QDS-Arrazola2,QDS-Amiri} and experimentally verified \cite{QDS-Clarke,QDS-Collins1,QDS-Collins2,QDS-Collins3,QDS-Yin2,QDS-Roberts,QDS-Zhang,QDS-AN}. To name a few, Ref. \cite{QDS-Clarke} and Ref. \cite{QDS-Collins1} reported the QDS demonstrations without the limitation of nondestructive state comparison and quantum memory, respectively, but still with secure quantum channels. Whereafter, Ref. \cite{QDS-Amiri} succeeded in reducing the technical complexity to the same level of QKD, thus removing the assumption of secure quantum channels. Naturally, decoy state method \cite{Decoy1,Decoy2} is adopted in QDS to resist photon-number-splitting attacks \cite{PNS1,PNS2}. All these developments of QDS have paved the way to its practical applications. 

In this paper, we focus on further improving its practical performance, i.e. higher signature rates and longer transmission distance with a simple experimental setup. Firstly, we apply the one-decoy method, which outperforms the two-decoy scheme for almost experimental settings in QKD \cite{one-decoy}. More importantly, this method in principle could decrease random number consumptions and experimental complexity in state preparation. Moreover, low-loss asymmetric Mach-Zehnder interfermeters (AMZIs) are employed, and, meanwhile, a real-time polarization calibration method is implemented to compensate for polarization drifts. As a result, we get an easy to operate and stable QDS system. With a rigorous statistical fluctuation analysis \cite{Hoeffding,M-Curty}, we attain a signature rate of 0.98 bit/s at 103 km, and reach the longest transmission distance of 280-km optical fibers so far.

Here, we consider a three-part scenario, including one signer, Alice, and two recipients, Bob and Charlie. Our one-decoy QDS consists of a distribution stage and a messaging stage. In distribution stage, Bob-Alice and Charlie-Alice individually utilize the one-decoy QKD protocol but without error correction and privacy amplification, namely key-generation protocol (KGP), to generate correlated bit strings. While in messaging stage, Alice transmits signature messages to the two recipients. The concrete procedure is outlined as follows:                                        

\textbf{\textit{Distribution stage.}} Bob (Charlie) prepares $N_t$ weak coherent state pulses, each of which is randomly modulated in $\varsigma  \in \{ X, Z\}$ basis and $\lambda  \in \left\{ {\mu, \nu } \right\}$ (signal, decoy) state, and sends them to Alice. Alice randomly selects $\varsigma  \in \{ X, Z\}$ basis to measure the received pulses. Later, Bob (Charlie) and Alice publicly announce their basis choices with an authentic channel and discard the mismatched cases. They agree that the data in \textit{Z} basis is to extract sifted keys and in \textit{X} basis to estimate parameters. After previous sifting procedure, they randomly reveal a small part of keys, say $k$-length, to estimate error rate $E_{BA}$ ($E_{CA}$), leaving the rest as a key pool for signature. To sign a message \textit{m = 0 or 1}, Alice and Bob (Charlie) choose \textit{L}-length block from key pool to construct signature sequence $K_m^B$ and $B_m^A$ ($K_m^C$ and $C_m^A$), where $K_m^B$ and $K_m^C$ is held by Alice, and $B_m^A$ ($C_m^A$) by Bob (Charlie). Finally, Bob (Charlie) randomly selects half of his keys and forwards them as well as corresponding positions to Charlie (Bob) with a secret classical channel. Define Bob's and Charlie's symmetrized keys as $S_m^B = (B_{m,keep}^A, C_{m,forward}^A)$ and $S_m^C = (C_{m,keep}^A, B_{m,forward}^A)$, respectively.

Through the error rate of test keys $E_{BA}$, the upper bound of observed error rate $E_{BA}^U$ can be estimated with the Serfling inequality \cite{Serfling}
\begin{equation}
E_{BA}^U = {E_{BA}} + \frac{2}{L}\sqrt {\frac{{(L/2 + 1)(L/2 + k)}}{{2k}}\ln \left( {1/{\epsilon _{PE}}} \right)} ,
\label{eq-1}
\end{equation}
where $\epsilon _{PE}$ denotes the failure probability of Eq. \ref{eq-1}. Similarly, the upper bound of observed error rate $E_{CA}^U$ can be estimated, and $E^U=max(E_{BA}^U,E_{CA}^U)$. As in Ref. \cite{QDS-Amiri}, the minimum rate $p_E$ at which Eve can introduce errors can be evaluated using 
\begin{equation}
h\left( {{p_E}} \right) = 2s_{Z,1}^L/L\left[ {1 - h\left( {\phi _{Z,1}^U} \right)} \right],
\label{eq-2}
\end{equation}  
where $h(x)=-xlog_2(x)-(1-x)log_2(1-x)$ is the binary Shannon entropy function; $s_{Z,1}^L$ and $\phi _{Z,1}^U$ denote the lower bound of single-photon counts and the upper bound of single-photon error rate, respectively, which can be estimated with one-decoy shceme \cite{one-decoy} in finite-size scenario. 

\textbf{\textit{Messaging stage.}} Alice sends $(m,Sig_m)$ to a recipient, say Bob, where $Sig_m = (K_m^B,K_m^C)$. Bob compares his keys $S_m^B$ with the received $(m,Sig_m)$ and records the number of mismatches. If the mismatches in both halves of $S_m^B$ are fewer than $s_\alpha L/2$, where $s_\alpha$ denotes the authentication threshold associated with the desired security of QDS protocol, Bob accepts the message and further forwards to Charlie; otherwise, Bob rejects the message and announces abortion of current protocol. Charlie checks the forwarded message in the same way as Bob, but using a different threshold $s_\upsilon$, with $0 < s_\alpha < s_\upsilon < 1/2$, to protect against repudiation by Alice. Charlie accepts the message if the mismatches are still fewer than $s_\upsilon L/2$ in both halves of his keys. 

Here, for simplicity, we set ${s_\alpha } = {E^U} + \frac{{{p_E} - {E^U}}}{3}$, and ${s_\upsilon } = {E^U} + \frac{{2\left( {{p_E} - {E^U}} \right)}}{3}$. For security analysis of QDS protocol, we should consider the probability of robustness, repudiation and forging, which are expressed as follws \cite{QDS-Amiri}:  
\begin{equation}
P\left( {{\rm{Robust}}} \right) \le 2{\epsilon _{PE}},
\label{eq-3}
\end{equation}
\begin{equation}
P\left( {{\rm{Repudiation}}} \right) \le 2\exp \left[ { - \frac{1}{4}{{\left( {{s_\upsilon } - {s_\alpha }} \right)}^2}L} \right],
\label{eq-4}
\end{equation}
\begin{equation}
P\left( {{\rm{Forge}}} \right) \le \alpha  + {\epsilon _F} + {10 \epsilon _{PE}},
\label{eq-5}
\end{equation}
where $\alpha$ and $\epsilon _F$ are associated with the probability of Bob finding a signature with an error rate smaller than $s_\upsilon$;
the addition of ${10 \epsilon _{PE}}$ denotes the failure probabilities of estimating channel parameters in finite-size scenario. $\epsilon _F$ is defined as 
\begin{equation}
{\epsilon _F} = \frac{1}{\alpha }\left[ {{2^{ - L/2\left\{ {2s_{Z,1}^L/L\left[ {1 - h\left( {\phi _{Z,1}^U} \right)} \right] - h\left( {{s_\upsilon }} \right)} \right\}}} + \epsilon } \right],
\label{eq-6}
\end{equation}
where $\epsilon$ is the failure probability of estimating Eve's information in accordance with smooth min-entropy. Finally, for the protocol to be secure, the system's security parameter $P_{sec} $ satisfies  
\begin{equation}
{P_{sec}} \ge \max \left\{ {P\left( {{\rm{Roubst}}} \right),P\left( {{\rm{Repudiation}}} \right),P\left( {{\rm{Forge}}} \right)} \right\}.
\label{eq-7}
\end{equation}

\begin{figure}
	\centering
	\includegraphics[width=\linewidth,height=4cm]{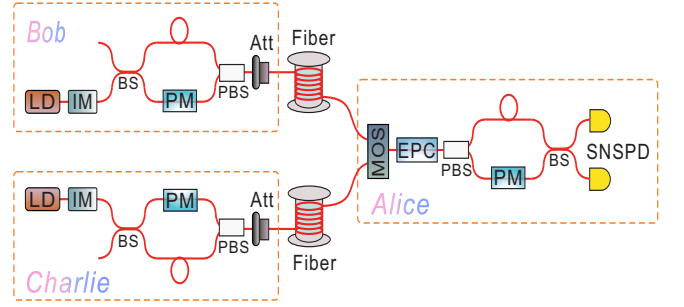}
	\caption{Schematic diagram of the experimental setup. LD, laser diode; IM, intensity modulator; BS, beam splitter; PM, phase modulator; PBS, polarization BS; Att, attenuator; MOS, mechanical optical switch; EPC, electronic polarization controller; SNSPD, superconducting nanowire single-photon detector. Low loss single-mode fibers are used in the link between Bob (Charlie) and Alice.}
	\label{fig-1}
\end{figure}

The optical layout of our experimental setup is sketched in Fig. \ref{fig-1}, where Bob and Charlie perform KGP with Alice separately in distribution stage. At transmitting Bob (Charlie) side, a laser diode (LD) pulsed at 50 MHz emits weak coherent pulses with 1550.12-nm center wavelength and 100-ps temporal duration. The laser possesses an intrinsic global phase randomization, which is a precondition to apply decoy-state technique. The intensity modulator (IM) is employed to modulate two different intensity levels, namely signal and decoy. Such one-decoy method omits the modulation of vacuum state, thus allowing single IM to provide a high dynamic extinction ratio and decreasing the modulation error. Moreover, it can reduce random number consumptions for controlling the IM. Quantum information is encoded upon optical phase using a homemade AMZI, with which each pulse is split into two divergent time bins (6.3 ns delay) of orthogonal polarizations. The AMZI consists of a beam splitter (BS), a phase modulator (PM) and a polarization BS (PBS), all of which are connected by polarization-maintaining fibers. In this way, we can reduce 3-dB loss in a pair of AMZIs, and thereby increase the useful counts. We should point it out that, limited by the modulation rate of PM, we adjust the phase at a low frequency, rather than completely random modulation. Then the coded pulses are weakened to single-photon level by an attenuator (Att) before transmitting to Alice through quantum channel. 

\begin{figure}
\centering
\includegraphics[width=8.4cm,height=4cm]{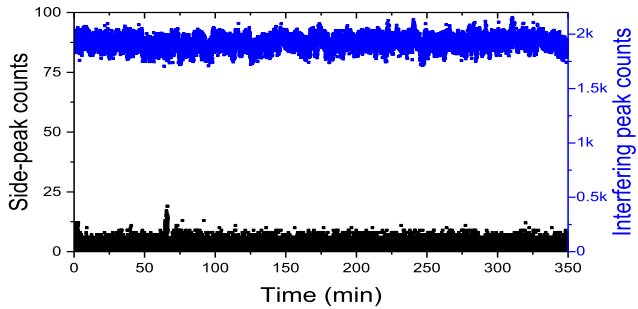}
\caption{Monitoring the side-peak and interfering peak stability under continuous operation of our QDS system for 6 hours at 204 km.}
\label{fig-2}
\end{figure}

At the measurement side, a mechanical optical switch (MOS) is inserted as routing function. Then, an electronic polarization controller (EPC) is deployed to adjust the polarization drifts introduced in installed fibers by monitoring and further suppressing the side-peak counts, with which our QDS system can keep  running continuously and stably without being interrupted. Fig. \ref{fig-2} shows the continuous operation of our QDS system at 204 km for 6 hours with an intensity of 0.32, where the blue (black) dot represents interfering (side-) peak counts. Alice decodes the phase information with another identical AMZI and directs the quantum signals into two commercial superconducting nanowire single-photon detectors (SNSPDs: TCOPRS-CCR-SW-85, SCONTEL company). The SNSPDs provide detection efficiencies about 65\% with dark count rates of 20 Hz when working at 2.3 K. The signals from detectors are collected with a time-to-digital converter (not shown in Fig. \ref{fig-1}), whose time window is set as 2 ns. The whole loss of measurement side is about 1.53 dB. Due to the effects of environmental fluctuations and drifts on optical interferometers, we adopt a scanning-and-transmitting mode \cite{Scan1,Scan2} to compensate phase drifts every 20 minutes. The resulting interference visibility is about 99.7\% and the duty cycle of transmission approximtes 86\%. In fact, we can also use an active phase compensation scheme \cite{LSTM} to further improve the system efficiency.

\begin{figure}
\centering
\includegraphics[width=8cm]{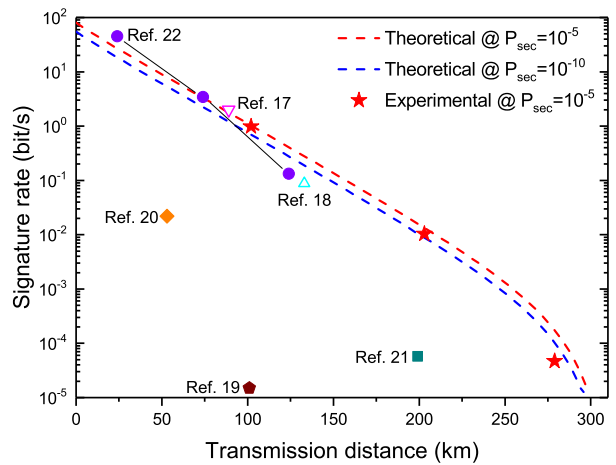}
\caption{The signature rates versus transmission distance. The upper (lower) line denotes our theoretical signature rates with the security parameter $10^{-5}$ ($10^{-10}$); while the asterisks denote corresponding experimental results at the transmission distance of 103-, 204- and 280-km installed fibers, respectively. In addition, other similar experimental reports are marked as different points for comparison.}
\label{fig-3}
\end{figure}

With the setup, we experimentally investigate the performance of one-decoy QDS at 103 km, 204 km and 280 km, respectively. For a better performance, all the intensities and probabilities are optimized according to following parameters: the fiber loss is measured as 0.175 dB/km; the optical misalignment error rate is approximately 0.3\%; and the total number of pulses sent by Bob or Charlie is $2\times 10^{12}$ at each distance. In addition, we reasonably set $\epsilon_{PE}=10^{-5}$, $\alpha=10^{-5}$ and $\epsilon=10^{-10}$. The theoretical and experimental results are shown in Fig. \ref{fig-3}, and the data of Refs. \cite{QDS-Collins2,QDS-Collins3,QDS-Yin2,QDS-Roberts,QDS-Zhang,QDS-AN} are also marked for comparison. The red dash line denotes the theoretical signature rates with the security parameter $P_{sec}=10^{-5}$; while the red asterisks corresponds to experimental signature rates at different distances. It can be seen that our experimental results agree well with theoretical predictions. The specific experimental values and estimated parameters are exhibited in Table \ref{table-1} and Table \ref{table-2}, respectively. For a fair comparison, we simulate another theoretical (blue dash) line with respect to $P_{sec}=10^{-10}$, which is a little worse than the red one. 

\begin{table*}
\renewcommand\arraystretch{1.25}
\centering
\caption{List of Total Counts ($n_\lambda ^\varsigma$) and Error Counts ($m_\lambda ^\varsigma$) at Different Distances. Numbers in Brackets Indicate Powers of 10.}
\setlength{\tabcolsep}{7.2mm}
\begin{tabular}{ccccccc}
\hline
\multirow{2}{*}{Event} &  \multicolumn{3}{c}{Bob-Alice} & \multicolumn{3}{c}{Charlie-Alice} \\
\cmidrule(r){2-4} \cmidrule(r){5-7}
& 103 km  & 204 km  & 280 km  & 103 km  & 204 km  & 280 km  \\ \hline
$n_\mu ^Z$ & 4.17[9] & 5.53[7] & 2.04[6] & 4.03[9] & 5.21[7] & 2.05[6] \\
$m_\mu ^Z$ & 7.35[6] & 165422  & 39090   & 6.66[6] & 171233  & 36028   \\
$n_\nu ^Z$ & 4.05[7] & 4.22[6] & 407033  & 4.09[7] & 4.00[6] & 396024  \\
$m_\nu ^Z$ & 77579   & 26223   & 21908   & 109820  & 25500   & 21838   \\
$n_\mu ^X$ & 1.84[6] & 253975  & 102804  & 1.85[6] & 254259  & 100303  \\
$m_\mu ^X$ & 1956    & 686     & 1831    & 2657    & 675     & 1854    \\
$n_\nu ^X$ & 19474   & 18389   & 20279   & 19240   & 19636   & 21236   \\
$m_\nu ^X$ & 68      & 104     & 1209    & 34      & 94      & 1086    \\
\hline
\end{tabular}
\label{table-1}
\end{table*}

\begin{table}
\centering
\caption{Overview of Estimated Parameters with Security Parameter of $10^{-5}$.}
\setlength{\tabcolsep}{1.5mm}
\begin{tabular}{cccccc}
\hline
Distance  & $s_{Z,1}^L$ & $\phi _{Z,1}^U$ & $s_\alpha$ & $s_\upsilon$ & Siganture Rate \\
\hline
103 km & 17250  & 2.18\% & 8.02\% & 10.81\% & 0.98 bit/s\\
204 km & 21877  & 2.53\% & 7.19\% & 9.59\%  & 0.01 bit/s\\
280 km & 139259 & 3.90\% & 4.67\% & 5.44\%   & $4.67\times 10^{-5}$ bit/s \\
\hline
\end{tabular}
\label{table-2}
\end{table}

As shown in Fig. \ref{fig-3}, the experimental results of Refs. \cite{QDS-Collins2,QDS-AN} are better than our theoretical signature rates at short distance, e.g., < 100 km, which is attributed mainly to their high repetition rate. The security of differential-phase-shift protocol used in Refs. \cite{QDS-Collins2,QDS-Collins3} against coherent attack has not been completely proven. Moreover, compared with other reports, we achieve better performance both in signature rates and transmission distance with a simple experimental system. The details of typical experimental parameters and corresponding time to sign a bit message are listed in Table \ref{table-3}. Among them, the KGP in Ref. \cite{QDS-Roberts} exploits measurement-device-independent QKD protocol, and therefore possesses the highest security level. Ref. \cite{QDS-Yin2} implements over a distance of 102 km based on SARG04 protocol. In their experimental system, the transmitter employs different laser diodes to prepare specific states. This practice might induce a side channel when pulse shape and pulse width vary for different laser diodes \cite{Huang}. As for Ref. \cite{QDS-Zhang}, it achieves the longest transmission distance to date, but a low signature rates. Besides, it adopts the intricate parametric down-conversion source, and a local detection that requires high performance of detectors.  

\begin{table*}
\begin{center}
\centering
\caption{Comparison of Parameters among Recent QDS Experiments.}
\setlength{\tabcolsep}{0.88mm}
\begin{tabular}{cccccccc}
\hline
& Ref. \cite{QDS-Collins2} & Ref. \cite{QDS-Collins3} & Ref. \cite{QDS-Yin2} & Ref. \cite{QDS-Roberts}$^\dagger$ & Ref. \cite{QDS-Zhang}$^\dagger$ & Ref. \cite{QDS-AN} & This Paper \\
\hline
Protocol & DPS & DPS & SARG04 & MDI & Passive BB84 & BB84 & BB84\\
Repetition rate & 1 GHz & 1 GHz & 75 MHz & 1 GHz & 76 MHz & 1 GHz & 50 MHz\\
Maximal distance & 90 km & 134.2 km & 102 km & 50 km & 200 km & 125 km & 280 km\\
Security parameter & $10^{-10}$ & $10^{-10}$ & $10^{-9}$ & $10^{-10}$ & $10^{-4}$ & $10^{-5}$ & $10^{-5}$\\
&  &  &  &  &  &  &  1.02 s @ 103 km \\
Signature time per bit  & about 1 s & 27.13 s & 66840 s & 45 s & 17391.3 s & 7.52 s & 96.52 s @ 204 km  \\
&  &  &  &  &  &  &  21407.36 s @ 280 km \\
\hline
\end{tabular}
\label{table-3}
\end{center}
{$\dagger$ With attenuator.}
\end{table*}

To conclude, we have implemented a practical QDS experiment using low loss AMZIs. It feature an efficient one-decoy scheme and real-time polarization calibration. Benefiting from above advantages, our QDS system is simple and stable. Compared with previous reports, under the security premise of unforgeability, nonrepudiation and transferability of QDS protocol, we successfully achieve higher signature rates and to date the longest transmission distance, which paves the way towards practical implementation of QDS in the near future.

We would like to acknowledge Jiangsu Hengtong Cable Co., LTD for low-loss single-mode fibers, and Dr. Shuang Wang for useful discussion on design of AMZIs. We also gratefully acknowledge the financial support from National Key Research and Development Program of China (2018YFA0306400, 2017YFA0304100); National Natural Science Foundation of China (NSFC) (11774180, 61590932, 61705110, 11847215); China Postdoctoral Science Foundation (2018M642281).

\end{document}